\pdfoutput=1

\documentclass[11pt]{article}

\usepackage[]{ACL2023}

\usepackage{times}
\usepackage{latexsym}
\usepackage{amsmath}
\usepackage{graphicx}
\usepackage{multirow}

\usepackage[T1]{fontenc}

\usepackage[utf8]{inputenc}

\usepackage{microtype}

\usepackage{inconsolata}

\title{Selective Shot Learning for Code Explanation}

\author{Paheli Bhattacharya \and Rishabh Gupta \\
        Bosch Research and Technology Centre, India\\
        \texttt{\{paheli.bhattacharya, gupta.rishabh\}@in.bosch.com}
        }

\begin{document}
\maketitle
\begin{abstract}
Code explanation plays a crucial role in the software engineering domain, aiding developers in grasping code functionality efficiently. 
Recent work shows that the performance of LLMs for code explanation improves in a few-shot setting, especially when the few-shot examples are selected intelligently. State-of-the-art approaches for such Selective Shot Learning (SSL) include token-based and embedding-based methods~\cite{geng}. However, these SSL approaches have been evaluated on proprietary LLMs, without much exploration on open-source Code-LLMs. Additionally, these methods lack consideration for programming language syntax. To bridge these gaps, we present a comparative study and propose a novel SSL method ($SSL_{ner}$) that utilizes entity information for few-shot example selection. We present several insights and show the effectiveness of $SSL_{ner}$ approach over state-of-the-art methods across two datasets. To the best of our knowledge, this is the first systematic benchmarking of open-source Code-LLMs while assessing the performances of the various few-shot examples selection approaches for the code explanation task.

\end{abstract}
\section{Introduction}
Code understanding and explanation~\cite{code_explain}, also known as code summarization~\cite{code_summ, code_summ_dl1} and code comment generation~\cite{code_comment_dl1, code_comment_lamner}, is an important problem in the domain of software engineering. It involves generating concise and informative explanations for pieces of source code. This provides the developers with a quick understanding of its functionality aiding in code maintenance, search and retrieval~\cite{ye2020leveraging}. For programmers new to a particular programming language, code summaries serve as valuable documentation to familiarize them with the new environment efficiently~\cite{code_explain}. Automating the task of code documentation through comments and explanations can therefore prove beneficial in many ways.

Large Language Models (LLMs) have proven their efficiency in a variety of NLP tasks. LLMs have shown promising results in several software engineering tasks like code generation~\cite{li2023python, yin-etal-2023-natural}, translation~\cite{huang2023program}, test case generation~\cite{schäfer2023empirical} and code explanation~\cite{geng, code_summ, code_explain, bhattacharya2023exploring, segment}. 
While using LLMs for the code explanation task,  It has been shown that few-shot prompting achieves better results than zero-shot prompting~\cite{geng, segment}. Hence, selecting the demonstration examples for few-shot learning is an important design criteria. We use the term Selective Shot Learning (SSL) when examples for few-shot prompting are chosen intelligently, instead of being random. SSL approaches for code explanation include token-based and embedding-based methods~\cite{geng} without taking into account the language syntax.

Recent work in the area of code explanation have only considered proprietary LLMs like Codex~\cite{geng,code_explain}, Code-davinci-002~\cite{code_summ}, Text-Davinci-003~\cite{segment}, GPT-3~\cite{code_explain} and GPT-3.5-turbo~\cite{segment}. However there is a huge gap in proper benchmarking and performance evaluation of several competing, open-source Code-LLMs like CodeLlama~\cite{codellama}, StarCoder~\cite{starcoder} for the code explanation task. 

\noindent
To this end, the contributions of the paper are:

\noindent
$\bullet$ We explore several open-source Code-LLMs for the task of code explanation, across two datasets covering different levels of code explanations (inline and method-level). To the best of our knowledge, this is the first systematic attempt at benchmarking the task. We make the dataset and code publicly available at \url{https://github.boschdevcloud.com/HXT2KOR/code-explanation}.

\noindent
$\bullet$ We assess the performance of several selective-shot learning approaches, including token-based and embedding-based approaches. Additionally we propose a novel Selective-shot Learning method using NER ($SSL_{ner}$) that includes code-based entity information for examples selection. 

\noindent
$\bullet$ We draw several interesting insights -- for e.g., we find that the performance of the medium-sized LLMs (StarCoder 15B) increase more rapidly compared to the larger-sized LLM (CodeLlama 34B) and $SSL_{ner}$ to be the best performing SSL approach and being interpretable.

\label{sec:intro}

\section{Related Work}
The \textbf{Code Explanation}~\cite{code_explain} task is a well studied problem in the domain of software engineering~\cite{trad1, trad3, trad2}. With the advent of deep learning, methods combining neural architectures~\cite{cai-etal-2020-tag, ahmad2020transformer, code_comment_lamner} along with software engineering approaches like AST trees~\cite{code_comment_dl1} have been proposed.

\textbf{Large Language Models} have shown exceptional performance in a plethora of NLG tasks~\cite{yang2023harnessing}. The zero-shot and few-shot capabilities of these model make them highly adaptable to many NLP tasks. Generic, open-source LLMs like LLama-2~\cite{touvron2023llama}, Alpaca~\cite{taori2023alpaca} are trained on open internet datasets. CodeLLMs such as StarCoder~\cite{starcoder}, CodeUp~\cite{codeup}, CodeLlama~\cite{codellama} and Llama-2-Coder~\cite{llama2coder} have been either trained or fine-tuned on code-specific datasets containing source codes covering around $80$+ programming languages.

The ~\textbf{Large Language Models}, when used for the ~\textbf{Code explanation} task, has shown some encouraging results. The recent approaches~\cite{code_explain, geng, code_summ, segment} demonstrate that the LLMs performs better in the few-shot setup when good examples of the task are provided. Hence, deciding the relevant examples is an important design criteria while using LLMs for the code explanation task. Existing approaches involve token-based, embedding-based~\cite{geng} and BM-25 along with repository information, data flow graph, AST tree etc.~\cite{segment}. However, these methods do not explore the efficacy of CodeLLMs. 

There has been \textbf{systematic evaluations} of transformer models (CodeT5 and CodeBERT) on code summarization~\cite{mondal2023understanding}, LLMs on code search~\cite{diera2023gencodesearchnet} and non-CodeLLMs like GPT, Bard for code documentation generation~\cite{10.1145/3664646.3664765}. To the best of our knowledge, this is the first work that performs an exhaustive analysis of four open-source CodeLLMs across two datasets and three SSL approaches for the task of code explanation. The SSL approaches explored in the paper do not incorporate secondary statistical analysis tools like AST tree, data-flow graph etc. like~\cite{segment}
\label{sec:rw}

\section{Dataset}
\begin{table*}[t]
\centering
\small
\caption{Statistics of the two datasets -- CoNaLa and TLC -- experimented within this paper. CoNaLa contains inline level codes written in Python. TLC contains function level codes written in Java. TLC is further subdivided into 5 different subdomains (code intents). CoNaLa contains shorter codes compared to TLC. The average length of the comments are comparable for the two datasets.}
\resizebox{\linewidth}{!}{
\label{tab:dataset}
\begin{tabular}{|c|c|c|c|cc|cccc|}
\hline
\multirow{3}{*}{\textbf{Code Level}} & \multirow{3}{*}{\textbf{Language}} & \multirow{3}{*}{\textbf{Dataset}} & \multirow{3}{*}{\textbf{Sub-domain}} & \multicolumn{2}{c|}{\textbf{\# Samples}} & \multicolumn{4}{c|}{\textbf{Average length}} \\ \cline{5-10} 
 &  &  &  & \multicolumn{1}{c|}{\multirow{2}{*}{train}} & \multirow{2}{*}{test} & \multicolumn{2}{c|}{train} & \multicolumn{2}{c|}{test} \\ \cline{7-10} 
 &  &  &  & \multicolumn{1}{c|}{} &  & \multicolumn{1}{c|}{Code} & \multicolumn{1}{c|}{Comment} & \multicolumn{1}{c|}{Code} & Comment \\ \hline
Inline & Python & CoNaLa & -- & \multicolumn{1}{c|}{1666} & 350 & \multicolumn{1}{c|}{13.92} & \multicolumn{1}{c|}{14.68} & \multicolumn{1}{c|}{14.35} & 14.06 \\ \hline
\multirow{6}{*}{Function} & \multirow{6}{*}{Java} & \multirow{6}{*}{TLC} & How-to-use & \multicolumn{1}{c|}{838} & 37 & \multicolumn{1}{c|}{75.14} & \multicolumn{1}{c|}{12.75} & \multicolumn{1}{c|}{65.41} & 12.97 \\ \cline{4-10} 
 &  &  & Property & \multicolumn{1}{c|}{5,016} & 292 & \multicolumn{1}{c|}{69.96} & \multicolumn{1}{c|}{12.86} & \multicolumn{1}{c|}{73.5} & 12.59 \\ \cline{4-10} 
 &  &  & Why & \multicolumn{1}{c|}{5,935} & 297 & \multicolumn{1}{c|}{82.29} & \multicolumn{1}{c|}{12.47} & \multicolumn{1}{c|}{83.38} & 12.34 \\ \cline{4-10} 
 &  &  & How-it-is-done & \multicolumn{1}{c|}{11,478} & 507 & \multicolumn{1}{c|}{89.5} & \multicolumn{1}{c|}{14.63} & \multicolumn{1}{c|}{89.94} & 14.32 \\ \cline{4-10} 
 &  &  & What & \multicolumn{1}{c|}{28,991} & 2158 & \multicolumn{1}{c|}{87.26} & \multicolumn{1}{c|}{11.8} & \multicolumn{1}{c|}{86.56} & 11.12 \\ \hline
\end{tabular}
}
\vspace{-5mm}
\end{table*}
In order to perform an extensive evaluation of the performance of the different open source Code-LLMs on the code explanation task, we consider two types of datasets  which have different levels of codes and explanations -- Inline level and Function level. We describe each of them in detail:\\
~\textbf{(i) Inline level: } This involves explaining particular lines of codes. Inline documentation improves readability and maintainability of a code. We experiment with the CoNaLa: The Code/Natural Language Challenge dataset~\cite{conala}. The dataset contains manually curated $(code\ snippet, code\ explanation)$ pairs. The code snippets are in the Python programming language. The code explanation is a natural language description that explains the task \textit{code snippet} is performing. Table~\ref{tab:dataset} shows the statistics of the dataset. There are  $1,666$ and $350$ samples in the train and test sets respectively. The average length of code snippet and their explanations is approximately $14$ tokens.\\
~\textbf{(ii) Function level: } This involves explaining specific functions or methods. We experiment with the TLC dataset~\cite{mu2023developer}, a widely-used dataset for the code comment generation task. The TLC dataset has additional labels for each data sample that implies the intents of the code -- ``how to use'', ``property'', ``why'', ``how it is done'' and ``what''. 
Since the code snippets in TLC dataset are function level codes, we find in Table~\ref{tab:dataset} that the length of the code snippets are longer than the ones in the CoNaLa dataset. However the length of the explanations is on average $12$ tokens which is comparable to CoNaLa. The test data size is 4,236 samples, with a minimum for the ``how-to-use'' intent (37 samples) and maximum for the ``what'' intent (2158 samples).

\label{sec:dataset}

\section{Selective-Shot Learning Approaches}
\begin{figure*}[t]
\caption{The workflow of the code explanation pipeline using Selective Shot Learning (SSL) approaches. In the input we have a query code snippet $q$ whose explanation needs to be generated and a training database containing $(code\ snippet, code\ explanation)$ pairs from which the few-shot examples need to be selected. The training data samples are ranked according to their similarity with $q$, where similarity can be computed using either $Selection_{token}$, $Selection_{semantic}$ or $SSL_{ner}$. From the ranked list, top-k examples are selected and given as a prompt along with $q$ to an LLM which then generates the explanation.}
\label{fig:workflow}
\centering
\includegraphics[width=16cm, height=8cm]{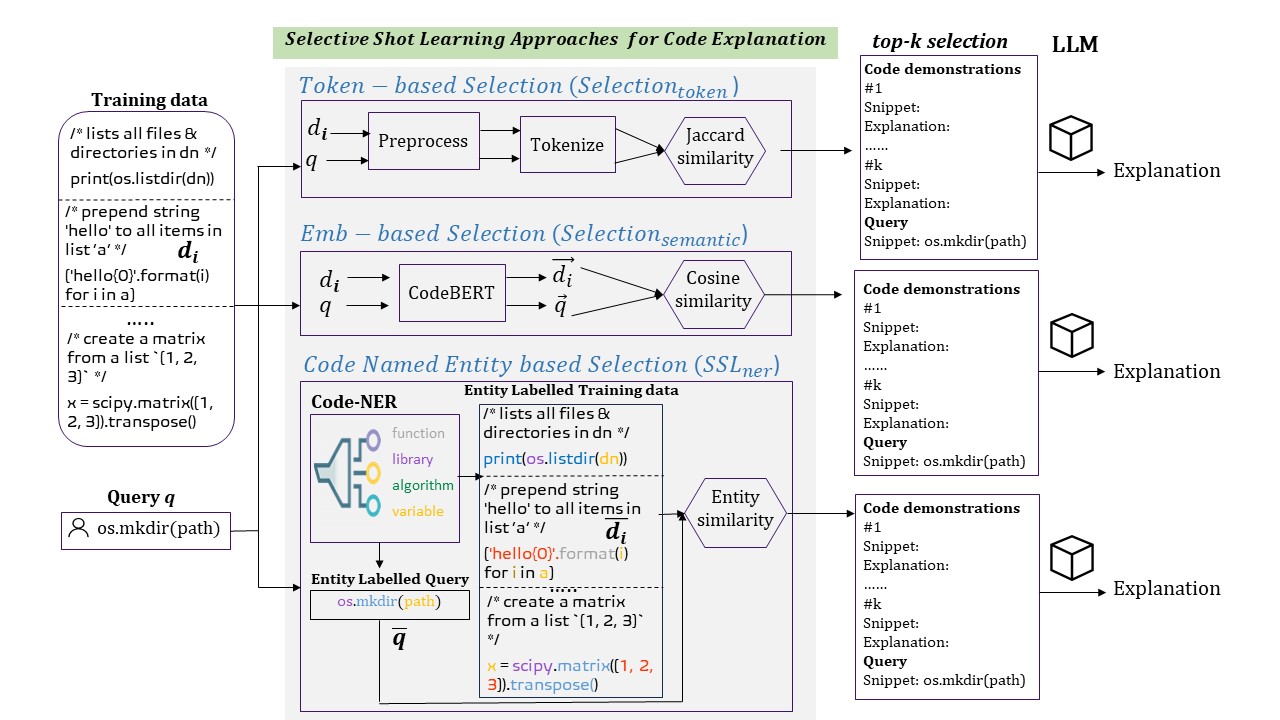}
\vspace{-5mm}
\end{figure*}

In this section we elaborate the different approaches for selecting relevant demonstrations for the code explanation task. The general pipeline is shown in  Figure~\ref{fig:workflow}. It is assumed that there is a database containing $(code\ snippet, code\ explanation)$ pairs (referred to as training data) from which relevant examples will be selected. Similarity is computed between the input code snippet ($q$) and all $code\ snippets$ ($d_i$) in the database, using the approaches $Selection_{token}$, $Selection_{semantic}$ and $SSL_{ner}$ described next. From each approach, we find the most relevant $k$ code snippets, along with their explanations, and curate a prompt which is then passed on to an LLM to generate the explanation for $q$.

\subsection{Token-based selection}
In the token-based selection strategy proposed in~\cite{geng} the query code $q$ and the all code snippets $d_i$ are first preprocessed by removing the keywords defined in the programming languages and converting all the tokens to lower case. The preprocessed $q$ and $d_{i}$'s are then converted to a list of tokens $tokens_{target}$ and $tokens_{candidate}$ respectively. Then a Jaccard similarity is computed between the two token lists to get the resulting token based similarity.

$Selection_{token}$ = $\frac{\left|\ tokens_{target}\  \cap \ tokens_{candidate} \ \right|}{\left| \ tokens_{target} \ \cup  \ tokens_{candidate}\  \right|}$. 
\noindent
The value of $Selection_{token}$ ranges from 0 to 1. A larger value of indicates a higher similarity between the query code and the candidate code from the retrieved set. Based on the similarity value, the $d_i$'s are ranked in decreasing order and then the top-k most similar code snippet and their corresponding explanation is added as few-shot demonstrations.

\subsection{Embedding-based selection}
\vspace{-2mm}
In the embedding-based approach proposed in~\cite{geng}, the query code $q$ and all code snippets $d_i$ in the database are encoded as vectors $\overrightarrow{d_i}$ and $\overrightarrow{q}$ respectively using the CodeBERT embedding model. The $Selection_{semantic}$ score is then the cosine similarity computed between the embeddings $\overrightarrow{d_i}$ and $\overrightarrow{q}$.  The value of $Selection_{semantic}$ lies between 0 to 1. A larger value indicates a higher similarity. Based on the similarity value, the $d_i$'s are ranked in decreasing order and then the top-k most similar code snippets and their corresponding explanations are added as few-shot demonstrations.

\subsection{Code Named Entity based Selection}
In this section, we present a novel method, Selective-shot Learning using Named Entity Recognition ($\boldsymbol{SSL_{ner}}$), that utilizes code-based named entities to select examples. It has two submodules Code Entity Extraction and Entity-based similarity, described subsequently.

\noindent
\textbf{Code entity extraction} -- This is the entity extraction module that returns a set of  entities $E$ from the programming language domain. We use UniversalNER~\cite{uniner}, an LLM that extracts entities from a wide variety of domains including programming. 20 different entities like function, library, data structure, algorithms etc. are supported in the model. For instance, given a code snippet \texttt{print(os.listdir(dname))}, this module will label \texttt{print} and \texttt{listdir} as `function', \texttt{os} as library and \texttt{dname} as `variable'. Figure~\ref{fig:workflow} shows that the training data samples and the query code are passed through the code entity extraction module and each of them are labelled with entity information. 

\noindent
\textbf{Entity-based similarity} -- This is the entity similarity module to find how similar are the list of entities which are extracted from the code snippets. Given two code snippets $q$ and $d$, the similarity:
\begin{equation}
\label{eq:ne}
    sim_{ne}(q,d) =\sum_{i=1}^{|E|} w_{e_{i}}*s_{e_{i}(q,d)}
    \vspace{-3mm}
\end{equation}

where $e_{i} \in E$ is a particular entity type; $s_{e_{i}}(q,d)  = jaccard(e_{i_{q}}, e_{i_{d}})$ is the jaccard similarity between $e_{i_{q}}, e_{i_{d}}$ (the entities of type $e_{i}$ in $q$ and $d$ respectively) and $w_{e_{i}}$ is the weightage for an entity type $e_{i}$ in similarity estimation. 
We assign $w_{e_{i}}=0$ for $e_{i}$ = `data type', `variable' and `value' because the entities of these types may not play a major role in similarity estimation. For others we set $w_{e_{i}}=1$.

To summarize, $\boldsymbol{SSL_{ner}}$ takes the input code snippet $q$ and the training database containing documented code pairs in the form of~$(code\ snippet, code\ explanation)$. These pairs are then ranked in decreasing order of similarity values $sim_{ne}(d,q)$ calculated using Eq.~\ref{eq:ne}. The top-$k$ most similar code snippets along with their explanations are selected, appended with the prompt and sent to an LLM to generate the explanation of the input code snippet $q$.

In the example (Figure~\ref{fig:workflow}), given a query code snippet \texttt{os.mkdir(path)} and $k=2$, the similar codes that are likely to get retrieved are \texttt{print(os.listdir(dname)} and \texttt{r+=[e for e in os.listdir(folder) if e.endswith(`.c')]}, since both these code snippets use the \texttt{os} library. The query code snippet \texttt{os.mkdir(path)} also uses the same library and hence is more similar to those two code snippets than others (e.g. \texttt{x=scipy.matrix([1,2,3]).transpose()}) in the training set. The code samples along with their explanations now forms the demonstrations in the prompt. 

\label{sec:app}

\section{Experimental Setup}
In this section we describe the experimental design choices used in this paper.

\textbf{Evaluation}
We use the BLEU, METEOR~\cite{geng} and ROUGE-L FScore for evaluating the model generated  explanations with respect to the ground truth explanations. These are the most widely used metrics for the task~\cite{geng, code_comment_dl1, segment}.


\textbf{Large Language Models}
We evaluate the performance of the different approaches by providing prompts to the following LLMs -- Llama-2-Coder-7B, CodeUp-13B-Chat, StarCoder (15.5B) and CodeLlama-34B-Instruct. We use $k=10$ examples as suggested by previous works~\cite{geng, code_summ} for better performance. For the UniversalNER LLM, we set max\_new\_tokens=64, do\_sample=False, temperature=0.1. For all CodeLLMs, we set max\_new\_tokens = 32, do\_sample = False and temperature = 0.7. 

For the TLC dataset, there are five intents as described in Section~\ref{sec:dataset}. ~\cite{geng} uses these intents in the prompt construction. For instance, for a test query code from the intent ``how-to-use'' they use the prompt: ``Describe the usage or the expected set-up of using the method''. However, we find that including such intent-specific keywords in the prompt does not affect the performance of the open source code LLMs. We therefore do not include the description of the intents in the prompt. 

The zero-shot prompt templates used in our experiments are as follows:

CodeLlama:
\texttt{
{[}INST{]} \textless \textgreater You are an expert in Programming. Below is a line of python code that describes a task. Return only one line of summary that appropriately describes the task that the code is performing. You must write only summary without any prefix or suffix explanations. Note: The summary should have minimum 1 words and can have on an average 10 words. \textless \textgreater \{\textit{code}\} {[}/INST{]}}

Llama2-Coder, StarCoder and CodeUp:\\
\texttt{
\#Human: You are a helpful code summarizer. Please describe in simple english the purpose of the following Python code snippet: \{\textit{code}\} \\\#Assistant:
}



\section{Results}
The empirical results of the code explanation task on the CoNaLa dataset are presented in Table~\ref{tab:conala_quant}. For the five code intents in the TLC dataset the results are given in Tables~\ref{tab:tlc_use_quant}--\ref{tab:tlc_what_quant}. We frame  research questions addressing the pivotal points in using LLMs for the task of code explanation and also the effects of different exemplar selection strategies. 

\vspace{2mm}
\noindent
\textbf{RQ1: The effectiveness of open-source CodeLLMs for the task of code explanation using the vanilla In-context learning technique.}
The first two rows in for each open source code LLM (LLama2-Coder, CodeUp, StarCoder and CodeLlama) in Tables ~\ref{tab:conala_quant}, ~\ref{tab:tlc_use_quant}--\ref{tab:tlc_what_quant} show the performance of zero-shot and randomly selected examples for few-shot prompting techniques (\textit{few shot (random)}).

\begin{table}[!thb]
\centering
\caption{The performance of the approaches using four LLMs for the code explanation task on the CoNaLa dataset. We report the \% improvement of $SSL_{ner}$ over the baseline approaches $Selection_{token}$ and $Selection_{semantic}$.}
\label{tab:conala_quant}
\resizebox{\linewidth}{!}{
\begin{tabular}{|c|c|c|c|c|}
\hline
\textbf{Model} & \textbf{Approach} & \textbf{BLEU} & \textbf{ROUGE-L} & \textbf{METEOR} \\ \hline
\multirow{7}{*}{\begin{tabular}[c]{@{}c@{}}Llama2-Coder\\ (7B)\end{tabular}} & zero shot & 0.292 & 0.298 & 0.236 \\ \cline{2-5} 
 & few shot (random) & 0.364 & 0.373 & 0.323 \\ \cline{2-5} 
 & $Selection_{token}$ & 0.393 & 0.401 & 0.36 \\ \cline{2-5} 
 & $Selection_{semantic}$ & 0.405 & 0.415 & 0.379 \\ \cline{2-5} 
 & $SSL_{ner}$ & \textbf{0.408} & \textbf{0.419} & \textbf{0.386} \\ \hline 
\multirow{7}{*}{\begin{tabular}[c]{@{}c@{}}CodeUp\\ (13B)\end{tabular}} & zero shot & 0.31 & 0.35 & 0.203 \\ \cline{2-5} 
 & few shot (random) & 0.345 & 0.372 & 0.291 \\ \cline{2-5} 
 & $Selection_{token}$ & 0.382 & 0.403 & 0.343 \\ \cline{2-5} 
 & $Selection_{semantic}$ & 0.402 & 0.417 & 0.368 \\ \cline{2-5} 
 & $SSL_{ner}$ & \textbf{0.412} & \textbf{0.424} & \textbf{0.384} \\ \hline 
\multirow{7}{*}{\begin{tabular}[c]{@{}c@{}}StarCoder\\ (15B)\end{tabular}} & zero shot & 0.291 & 0.33 & 0.216 \\ \cline{2-5} 
 & few shot (random) & 0.373 & 0.402 & 0.335 \\ \cline{2-5} 
 & $Selection_{token}$ & 0.411 & 0.435 & 0.385 \\ \cline{2-5} 
 & $Selection_{semantic}$ & 0.429 & 0.449 & 0.407 \\ \cline{2-5} 
 & $SSL_{ner}$ & \textbf{0.435} & \textbf{0.451} & \textbf{0.416} \\ \hline 
\multirow{7}{*}{\begin{tabular}[c]{@{}c@{}}CodeLlama\\ (34B)\end{tabular}} & zero shot & 0.354 & 0.374 & 0.254 \\ \cline{2-5} 
 & few shot (random) & 0.369 & 0.38 & 0.321 \\ \cline{2-5} 
 & $Selection_{token}$ & 0.389 & 0.397 & 0.357 \\ \cline{2-5} 
 & $Selection_{semantic}$ & 0.395 & 0.403 & 0.375 \\ \cline{2-5} 
 & $SSL_{ner}$ & \textbf{0.399} & \textbf{0.405} & \textbf{0.381} \\ \hline
\end{tabular}
}
\vspace{-5mm}
\end{table}

\begin{table}[!thb]
\caption{The performance of all the approaches using four LLMs for the code explanation task over the \textbf{How-to-use} intent in the TLC dataset. We report the \% improvement of $SSL_{ner}$ over the baseline approaches $Selection_{token}$ and $Selection_{semantic}$.}
\label{tab:tlc_use_quant}
\resizebox{\linewidth}{!}{
\begin{tabular}{|c|c|c|c|c|}
\hline
\multirow{2}{*}{\textbf{Model}} & \multirow{2}{*}{\textbf{Approach}} & \multirow{2}{*}{\textbf{BLEU}} & \multirow{2}{*}{\textbf{ROUGE-L}} & \multirow{2}{*}{\textbf{METEOR}} \\
 &  &  &  &  \\ \hline
\multirow{7}{*}{\begin{tabular}[c]{@{}c@{}}Llama2-Coder\\ (7B)\end{tabular}} & zero shot & \multicolumn{1}{c|}{0.186} & \multicolumn{1}{c|}{0.126} & 0.123 \\ \cline{2-5} 
 & few shot (random) & \multicolumn{1}{c|}{0.291} & \multicolumn{1}{c|}{0.275} & 0.236 \\ \cline{2-5} 
 & $Selection_{token}$ & \multicolumn{1}{c|}{0.324} & \multicolumn{1}{c|}{0.315} & 0.291 \\ \cline{2-5} 
 & $Selection_{semantic}$ & \multicolumn{1}{c|}{0.347} & \multicolumn{1}{c|}{0.34} & 0.317 \\ \cline{2-5} 
 & $SSL_{ner}$ & \multicolumn{1}{c|}{\textbf{0.358}} & \multicolumn{1}{c|}{\textbf{0.355}} & \textbf{0.323} \\ \hline 
\multirow{7}{*}{\begin{tabular}[c]{@{}c@{}}CodeUp\\ (13B)\end{tabular}} & zero shot & \multicolumn{1}{c|}{0.187} & \multicolumn{1}{c|}{0.132} & 0.15 \\ \cline{2-5} 
 & few shot (random) & \multicolumn{1}{c|}{0.319} & \multicolumn{1}{c|}{0.302} & 0.274 \\ \cline{2-5} 
 & $Selection_{token}$ & \multicolumn{1}{c|}{0.342} & \multicolumn{1}{c|}{0.357} & 0.336 \\ \cline{2-5} 
 & $Selection_{semantic}$ & \multicolumn{1}{c|}{0.391} & \multicolumn{1}{c|}{0.381} & 0.367 \\ \cline{2-5} 
 & $SSL_{ner}$ & \multicolumn{1}{c|}{\textbf{0.395}} & \multicolumn{1}{c|}{\textbf{0.395}} & \textbf{0.372} \\ \hline 
\multirow{7}{*}{\begin{tabular}[c]{@{}c@{}}StarCoder\\ (15.5B)\end{tabular}} & zero shot & \multicolumn{1}{c|}{0.194} & \multicolumn{1}{c|}{0.138} & 0.107 \\ \cline{2-5} 
 & few shot (random) & \multicolumn{1}{c|}{0.259} & \multicolumn{1}{c|}{0.265} & 0.216 \\ \cline{2-5} 
 & $Selection_{token}$ & \multicolumn{1}{c|}{0.365} & \multicolumn{1}{c|}{0.393} & 0.351 \\ \cline{2-5} 
 & $Selection_{semantic}$ & \multicolumn{1}{c|}{0.402} & \multicolumn{1}{c|}{0.426} & 0.371 \\ \cline{2-5} 
 & $SSL_{ner}$ & \multicolumn{1}{c|}{\textbf{0.411}} & \multicolumn{1}{c|}{\textbf{0.431}} & \textbf{0.378} \\ \hline 
\multirow{7}{*}{\begin{tabular}[c]{@{}c@{}}CodeLlama\\ (34B)\end{tabular}} & zero shot & \multicolumn{1}{c|}{0.198} & \multicolumn{1}{c|}{0.136} & 0.173 \\ \cline{2-5} 
 & few shot (random) & \multicolumn{1}{c|}{0.237} & \multicolumn{1}{c|}{0.229} & 0.196 \\ \cline{2-5} 
 & $Selection_{token}$ & \multicolumn{1}{c|}{0.242} & \multicolumn{1}{c|}{0.206} & 0.263 \\ \cline{2-5} 
 & $Selection_{semantic}$ & \multicolumn{1}{c|}{0.263} & \multicolumn{1}{c|}{0.219} & 0.285 \\ \cline{2-5} 
 & $SSL_{ner}$ & \multicolumn{1}{c|}{\textbf{0.27}} & \multicolumn{1}{c|}{\textbf{0.223}} & \textbf{0.292} \\ \hline
\end{tabular}
}
\vspace{-5mm}
\end{table}

\begin{table}[!thb]
\caption{The performance of all the approaches using four LLMs for the code explanation task over the \textbf{why} intent in the TLC dataset. We report the \% improvement of $SSL_{ner}$ over the baseline approaches $Selection_{token}$ and $Selection_{semantic}$.}
\label{tab:tlc_why_quant}
\resizebox{\linewidth}{!}{
\begin{tabular}{|c|c|c|c|c|}
\hline
\multirow{2}{*}{\textbf{Model}} & \multirow{2}{*}{\textbf{Approach}} & \multirow{2}{*}{\textbf{BLEU}} & \multirow{2}{*}{\textbf{ROUGE-L}} & \multirow{2}{*}{\textbf{METEOR}} \\
 &  &  &  &  \\ \hline
\multirow{7}{*}{\begin{tabular}[c]{@{}c@{}}Llama2-Coder\\ (7B)\end{tabular}} & zero shot & 0.201 & 0.142 & 0.118 \\ \cline{2-5} 
 & few shot (random) & 0.261 & 0.221 & 0.196 \\ \cline{2-5} 
 & $Selection_{token}$ & 0.304 & 0.287 & 0.264 \\ \cline{2-5} 
 & $Selection_{semantic}$ & 0.346 & 0.318 & 0.288 \\ \cline{2-5} 
 & $SSL_{ner}$ & \textbf{0.352} & \textbf{0.324} & \textbf{0.298} \\ \hline
\multirow{7}{*}{\begin{tabular}[c]{@{}c@{}}CodeUp\\ (13B)\end{tabular}} & zero shot & 0.212 & 0.129 & 0.16 \\ \cline{2-5} 
 & few shot (random) & 0.257 & 0.231 & 0.21 \\ \cline{2-5} 
 & $Selection_{token}$ & 0.276 & 0.262 & 0.244 \\ \cline{2-5} 
 & $Selection_{semantic}$ & 0.296 & 0.289 & 0.268 \\ \cline{2-5} 
 & $SSL_{ner}$ & \textbf{0.301} & \textbf{0.297} & \textbf{0.276} \\ \cline{2-5} 
 & Gain (\%)   over $Selection_{token}$ & 9.06 & 13.36 & 13.11 \\ \cline{2-5} 
 & Gain (\%)   over $Selection_{semantic}$ & 1.69 & 2.77 & 2.99 \\ \hline
\multirow{7}{*}{\begin{tabular}[c]{@{}c@{}}StarCoder\\ (15.5B)\end{tabular}} & zero shot & 0.196 & 0.159 & 0.127 \\ \cline{2-5} 
 & few shot (random) & 0.278 & 0.279 & 0.242 \\ \cline{2-5} 
 & $Selection_{token}$ & 0.296 & 0.313 & 0.268 \\ \cline{2-5} 
 & $Selection_{semantic}$ & 0.315 & 0.331 & 0.297 \\ \cline{2-5} 
 & $SSL_{ner}$ & \textbf{0.338} & \textbf{0.342} & \textbf{0.303} \\ \hline
\multirow{7}{*}{\begin{tabular}[c]{@{}c@{}}CodeLlama\\ (34B)\end{tabular}} & zero shot & 0.225 & 0.186 & 0.216 \\ \cline{2-5} 
 & few shot (random) & 0.253 & 0.191 & 0.238 \\ \cline{2-5} 
 & $Selection_{token}$ & 0.313 & 0.294 & 0.315 \\ \cline{2-5} 
 & $Selection_{semantic}$ & 0.348 & 0.338 & 0.343 \\ \cline{2-5} 
 & $SSL_{ner}$ & \textbf{0.361} & \textbf{0.344} & \textbf{0.35} \\ \hline
\end{tabular}
}
\vspace{-3mm}
\end{table}

\begin{table}[!thb]
\caption{The performance of all the approaches using four LLMs for the code explanation task over the \textbf{property} intent in the TLC dataset. We report the \% improvement of $SSL_{ner}$ over the baseline approaches $Selection_{token}$ and $Selection_{semantic}$.}
\label{tab:tlc_property_quant}
\resizebox{\linewidth}{!}{
\begin{tabular}{|c|c|c|c|c|}
\hline
\textbf{Model} & \textbf{Approach} & \textbf{BLEU} & \textbf{ROUGE-L} & \textbf{METEOR} \\ \hline
\multirow{7}{*}{\begin{tabular}[c]{@{}c@{}}Llama2-Coder\\ (7B)\end{tabular}} & zero shot & 0.245 & 0.226 & 0.197 \\ \cline{2-5} 
 & few shot (random) & 0.323 & 0.341 & 0.305 \\ \cline{2-5} 
 & $Selection_{token}$ & 0.356 & 0.362 & 0.324 \\ \cline{2-5} 
 & $Selection_{semantic}$ & 0.391 & 0.405 & 0.359 \\ \cline{2-5} 
 & $SSL_{ner}$ & \textbf{0.401} & \textbf{0.416} & \textbf{0.372} \\ \hline
\multirow{7}{*}{\begin{tabular}[c]{@{}c@{}}CodeUp\\ (13B)\end{tabular}} & zero shot & 0.263 & 0.202 & 0.22 \\ \cline{2-5} 
 & few shot (random) & 0.429 & 0.42 & 0.404 \\ \cline{2-5} 
 & $Selection_{token}$ & 0.469 & 0.491 & 0.474 \\ \cline{2-5} 
 & $Selection_{semantic}$ & 0.528 & 0.517 & 0.505 \\ \cline{2-5} 
 & $SSL_{ner}$ & \textbf{0.542} & \textbf{0.532} & \textbf{0.522} \\ \hline
\multirow{7}{*}{\begin{tabular}[c]{@{}c@{}}StarCoder\\ (15.5B)\end{tabular}} & zero shot & 0.269 & 0.243 & 0.223 \\ \cline{2-5} 
 & few shot (random) & 0.456 & 0.476 & 0.446 \\ \cline{2-5} 
 & $Selection_{token}$ & 0.467 & 0.479 & 0.474 \\ \cline{2-5} 
 & $Selection_{semantic}$ & 0.544 & 0.524 & 0.531 \\ \cline{2-5} 
 & $SSL_{ner}$ & \textbf{0.558} & \textbf{0.535} & \textbf{0.538} \\ \hline
\multirow{7}{*}{\begin{tabular}[c]{@{}c@{}}CodeLlama\\ (34B)\end{tabular}} & zero shot & 0.252 & 0.215 & 0.254 \\ \cline{2-5} 
 & few shot (random) & 0.3 & 0.246 & 0.267 \\ \cline{2-5} 
 & $Selection_{token}$ & 0.337 & 0.328 & 0.377 \\ \cline{2-5} 
 & $Selection_{semantic}$ & 0.376 & 0.375 & 0.427 \\ \cline{2-5} 
 & $SSL_{ner}$ & \textbf{0.379} & \textbf{0.382} & \textbf{0.432} \\ \hline
\end{tabular}
}
\vspace{-3mm}
\end{table}

\begin{table}[!thb]
\caption{The performance of all the approaches using four LLMs for the code explanation task over the \textbf{How-it-is-done} intent in the TLC dataset. We report the \% improvement of $SSL_{ner}$ over the baseline approaches $Selection_{token}$ and $Selection_{semantic}$.}
\label{tab:tlc_done_quant}
\resizebox{\linewidth}{!}{
\begin{tabular}{|c|c|c|c|c|}
\hline
\textbf{Model} & \textbf{Approach} & \textbf{BLEU} & \textbf{ROUGE-L} & \textbf{METEOR} \\ \hline
\multirow{7}{*}{\begin{tabular}[c]{@{}c@{}}Llama2-Coder\\ (7B)\end{tabular}} & zero shot & 0.187 & 0.193 & 0.157 \\ \cline{2-5} 
 & few shot (random) & 0.271 & 0.267 & 0.235 \\ \cline{2-5} 
 & $Selection_{token}$ & 0.324 & 0.342 & 0.318 \\ \cline{2-5} 
 & $Selection_{semantic}$ & 0.357 & 0.372 & 0.348 \\ \cline{2-5} 
 & $SSL_{ner}$ & \textbf{0.366} & \textbf{0.387} & \textbf{0.358} \\ \hline
\multirow{7}{*}{\begin{tabular}[c]{@{}c@{}}CodeUp\\ (13B)\end{tabular}} & zero shot & 0.204 & 0.185 & 0.181 \\ \cline{2-5} 
 & few shot (random) & 0.292 & 0.297 & 0.259 \\ \cline{2-5} 
 & $Selection_{token}$ & 0.32 & 0.336 & 0.294 \\ \cline{2-5} 
 & $Selection_{semantic}$ & 0.36 & 0.366 & 0.325 \\ \cline{2-5} 
 & $SSL_{ner}$ & \textbf{0.369} & \textbf{0.371} & \textbf{0.327} \\ \hline
\multirow{7}{*}{\begin{tabular}[c]{@{}c@{}}StarCoder\\ (15.5B)\end{tabular}} & zero shot & 0.243 & 0.193 & 0.146 \\ \cline{2-5} 
 & few shot (random) & 0.331 & 0.338 & 0.327 \\ \cline{2-5} 
 & $Selection_{token}$ & 0.411 & 0.437 & 0.394 \\ \cline{2-5} 
 & $Selection_{semantic}$ & 0.449 & 0.486 & 0.427 \\ \cline{2-5} 
 & $SSL_{ner}$ & \textbf{0.463} & \textbf{0.491} & \textbf{0.436} \\ \hline
\multirow{7}{*}{\begin{tabular}[c]{@{}c@{}}CodeLlama\\ (34B)\end{tabular}} & zero shot & 0.262 & 0.211 & 0.232 \\ \cline{2-5} 
 & few shot (random) & 0.275 & 0.241 & 0.257 \\ \cline{2-5} 
 & $Selection_{token}$ & 0.325 & 0.325 & 0.309 \\ \cline{2-5} 
 & $Selection_{semantic}$ & 0.365 & 0.357 & 0.354 \\ \cline{2-5} 
 & $SSL_{ner}$ & \textbf{0.373} & \textbf{0.367} & \textbf{0.368} \\ \hline
\end{tabular}
}
\vspace{-3mm}
\end{table}

\begin{table}[!thb]
\caption{The performance of all the approaches using four LLMs for the code explanation task over the \textbf{What} intent in the TLC dataset. We report the \% improvement of $SSL_{ner}$ over the baseline approaches $Selection_{token}$ and $Selection_{semantic}$.}
\label{tab:tlc_what_quant}
\resizebox{\linewidth}{!}{
\begin{tabular}{|c|c|c|c|c|}
\hline
\textbf{Model} & \textbf{Approach} & \textbf{BLEU} & \textbf{ROUGE-L} & \textbf{METEOR} \\ \hline
\multirow{7}{*}{\begin{tabular}[c]{@{}c@{}}Llama2-Coder\\ (7B)\end{tabular}} & zero shot & 0.153 & 0.162 & 0.128 \\ \cline{2-5} 
 & few shot (random) & 0.285 & 0.274 & 0.242 \\ \cline{2-5} 
 & $Selection_{token}$ & 0.334 & 0.342 & 0.306 \\ \cline{2-5} 
 & $Selection_{semantic}$ & 0.352 & 0.358 & 0.317 \\ \cline{2-5} 
 & $SSL_{ner}$ & \textbf{0.358} & \textbf{0.363} & \textbf{0.325} \\ \hline
\multirow{7}{*}{\begin{tabular}[c]{@{}c@{}}CodeUp\\ (13B)\end{tabular}} & zero shot & 0.178 & 0.162 & 0.221 \\ \cline{2-5} 
 & few shot (random) & 0.312 & 0.41 & 0.368 \\ \cline{2-5} 
 & $Selection_{token}$ & 0.352 & 0.382 & 0.352 \\ \cline{2-5} 
 & $Selection_{semantic}$ & 0.392 & 0.41 & 0.373 \\ \cline{2-5} 
 & $SSL_{ner}$ & \textbf{0.407} & \textbf{0.425} & \textbf{0.381} \\ \hline
\multirow{7}{*}{\begin{tabular}[c]{@{}c@{}}StarCoder\\ (15.5B)\end{tabular}} & zero shot & 0.2 & 0.18 & 0.131 \\ \cline{2-5} 
 & few shot (random) & 0.291 & 0.327 & 0.274 \\ \cline{2-5} 
 & $Selection_{token}$ & 0.327 & 0.395 & 0.317 \\ \cline{2-5} 
 & $Selection_{semantic}$ & 0.365 & 0.403 & 0.354 \\ \cline{2-5} 
 & $SSL_{ner}$ & \textbf{0.374} & \textbf{0.416} & \textbf{0.362} \\ \hline
\multirow{7}{*}{\begin{tabular}[c]{@{}c@{}}CodeLlama\\ (34B)\end{tabular}} & zero shot & 0.193 & 0.183 & 0.234 \\ \cline{2-5} 
 & few shot (random) & 0.203 & 0.216 & 0.27 \\ \cline{2-5} 
 & $Selection_{token}$ & 0.28 & 0.287 & 0.287 \\ \cline{2-5} 
 & $Selection_{semantic}$ & 0.301 & 0.316 & 0.335 \\ \cline{2-5} 
 & $SSL_{ner}$ & \textbf{0.318} & \textbf{0.322} & \textbf{0.341} \\ \hline
\end{tabular}
}
\vspace{-3mm}
\end{table}

In both the CoNaLa and TLC datasets we observe CodeLlama to perform the best in the zero shot prompting setting. This is because the model is the largest in size (34B) compared to other models Llama2-Coder (7B), CodeUp (13B) and StarCoder (15.5B). Additionally, CodeLlama is further finetuned on Llama-2 while CodeUp and StarCoder has been trained for scratch on code data.

Interestingly, for the few shot prompting, we observe that the improvements over the zero-shot strategy are much more profound in the smaller sized models (Llama2-Coder, CodeUp and StarCoder) compared to CodeLlama. For instance, one can note from Table~\ref{tab:tlc_why_quant} that while CodeLlama $(0.225, 0.186, 0.216)$ performs better than StarCoder $(0.196, 0.159, 0.127)$ in the zero shot setting, the latter outperforms the former in the few shot setting, i.e., StarCoder in random few-shot gives $(0.278, 0.279, 0.242)$ and CodeLlama gives $(0.253, 0.191, 0.238)$. This could be attributed to the fact that since CodeLlama is a larger model, in-context examples does not add much to its existing, inherent knowledge. Smaller size models benefit further by providing in-context examples.
\begin{table*}[t]
\centering
\caption{An example showing a code snippet, its ground truth explanation, top 3 examples selected from the baseline method ($Selection_{token}$) and our approach ($SSL_{ner}$) and  the corresponding outputs generated by CodeLlama.}
\label{tab:qual}
\resizebox{\linewidth}{!}{
\begin{tabular}{|c|c|cl|cl|}
\hline
\multirow{2}{*}{\textbf{Snippet}} & \multirow{2}{*}{\textbf{Ground Truth}} & \multicolumn{2}{c|}{\textbf{$\boldsymbol{Selection_{token}}$}} & \multicolumn{2}{c|}{\textbf{$\boldsymbol{SSL_{ner}}$}} \\ \cline{3-6} 
 &  & \multicolumn{1}{c|}{\textbf{Top 3 Examples}} & \multicolumn{1}{c|}{\textbf{LLM Output}} & \multicolumn{1}{c|}{\textbf{Top 3 Examples}} & \multicolumn{1}{c|}{\textbf{LLM Output}} \\ \hline
\begin{tabular}[c]{@{}l@{}}open(filename, 'w').\\ close()\end{tabular} & \begin{tabular}[c]{@{}l@{}}erase all the contents\\ of a file`filename`\end{tabular} & \multicolumn{1}{l|}{\begin{tabular}[c]{@{}l@{}}\underline{snippet}: open('filename',   'w').write\\ ('\textbackslash{}\textbackslash{}n'.join('\%s \%s' \% x for x in mylist))\\ \underline{comment}: create a file 'filename'\\ with each tuple in the list `mylist`\\ written to a line\\      \\ \underline{snippet}: hashlib.md5(open('filename.exe',\\ 'rb').read()).hexdigest()\\ \underline{comment}: calculate the md5 checksum\\ of a file named  'filename.exe'\\  \\ \underline{snippet}: {[}w for w in open('file.txt') if not\\ re.search('{[}aeiou{]}\{2\}',   w){]}\\ \underline{comment}: Create a list containing \\ words that contain vowel letter\\ followed by the same vowel in file 'file.text'\end{tabular}} & \begin{tabular}[c]{@{}l@{}}The code creates a\\ new file with the name\\ `filename` and\\ closes it.\\\end{tabular} & \multicolumn{1}{l|}{\begin{tabular}[c]{@{}l@{}}\underline{snippet}: f =   open('test/test.pdf', 'rb')\\ \underline{comment}: read a binary file 'test/test.pdf\\      \\ \\ \underline{snippet}: f = open(os.path.join(\\ \_\_location\_\_, 'bundled-resource.jpg'))\\ \underline{comment}: open a file 'bundled-resource.\\ jpg'  in the same  directory as a\\ python   script\\      \\ \\ \underline{snippet}: f = open('words.txt', 'rU')\\ \underline{comment}: open the file 'words.txt'\\ in 'rU' mode\end{tabular}} & \begin{tabular}[c]{@{}l@{}}The code is erasing\\ the contents of a\\ file `filename`.\end{tabular} \\ \hline
\end{tabular}
}
\vspace{-5mm}
\end{table*}

\noindent
\textbf{RQ2 : Does the performance of open-source Code LLMs improve when provided with relevant in-context examples?}

Given that few shot learning improves performance over zero-shot, we now analyse if the quality of the few-shot examples affect the quality of the code explanations generated. As described in Section~\ref{sec:app}, we experiment with three few-shot example selection strategies -- $Selection_{token}$, $Selection_{semantic}$ and $SSL_{ner}$. 

For both the datasets, we observe that token and embedding-based demonstration selection strategies improve the code explanation generation performance across all the LLMs.  For instance in the CoNaLa dataset (Table~\ref{tab:conala_quant}) the BLEU scores for LLama2-Coder, CodeUp, StarCoder and CodeLlama increase by $12\%, 19\%, 17\%$ and $8\%$ respectively when compared with random few shot and $SSL_{ner}$. Similar to what we observed above, the improvements are more pronounced in the medium sized models, CodeUp and StarCoder, as compared to CodeLlama which is a 34B model. For the TLC dataset we observe this trend for intents ``how-to-use'', ``property'' and ``what'' (Tables~\ref{tab:tlc_use_quant}, ~\ref{tab:tlc_property_quant}, ~\ref{tab:tlc_what_quant}).

\vspace{2mm}
\noindent
\textbf{RQ3 : How do the token-based demonstration selection strategies compare?}

We now analyse the two token based demonstration selection strategies $Selection_{token}$ and $SSL_{ner}$. 

For CoNaLa dataset (Table~\ref{tab:conala_quant}), we find that $SSL_{ner}$ shows a better performance as compared to $Selection_{token}$. For instance, in the BLEU metric the improvements reported are $3.8\%, 7.85\%, 5.84\%$ and $2.57\%$ respectively for Llama2-Coder, CodeUp, StarCoder and CodeLlama. The improvements are statistically significant as measured paired Student's T-test at 95\%. 

Table~\ref{tab:qual} shows an example code snippet from the CoNaLa dataset, its ground truth explanation, the top 3 examples selected using $Selection_{token}$ and $SSL_{ner}$ and the corresponding outputs generated by the LLM model CodeLlama. The main intent of the example code snippet is to `erase' the contents of a file. The explanation generated by the $SSL_{ner}$ example selection strategy is more similar to the ground truth than the one by $Selection_{token}$. The examples selected by $SSL_{ner}$ are more concretely on `file opening' alone but $Selection_{token}$ selects examples that although have a notion of `opening the file' but is followed by subsequent, complex actions like calculating the checksum, performing string operations etc. This is likely to confuse the model thereby providing an erroneous explanation.



In the TLC dataset, we find that the improvements of $SSL_{ner}$ over $Selection_{token}$ are more notable. For instance, the gain \% achieved by $SSL_{ner}$ over $Selection_{token}$ for the intent ``what'' (which has the highest number of test samples, 2158, ref. Table~\ref{tab:dataset}) using CodeLlama and StarCoder in BLEU, ROUGE and METEOR are $(13\%,13.5\%,13.9\%)$ and $(14.6\%,9.6\%,11.82\%)$ respectively. These improvements are statistically significant.
\begin{figure*}[t]
\caption{An example demonstrating the Query Code method, the top 1 demonstration example selected by $Selection_{token}$, $Selection_{semantic}$ and $SSL_{ner}$ along with the LLM (StarCoder) generated output for each method, respectively.}
\label{fig:compare}
\centering
\fbox{
\includegraphics[width=16cm]{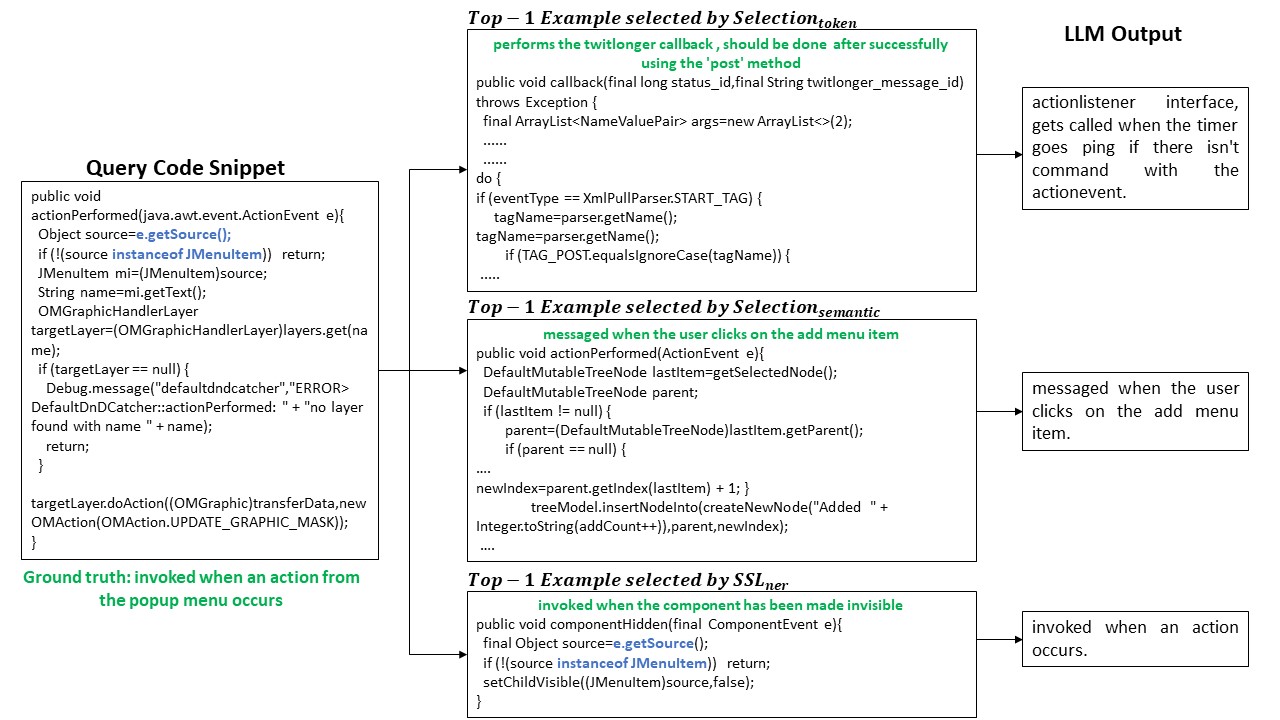}
}
\vspace{-7mm}
\end{figure*}

Hence we  conclude that $SSL_{ner}$ selects more relevant and consise demonstrations compared to the simpler $Selection_{token}$ approach (ref.~\ref{sec:app2} for more examples.). The method is interpretable through the matches in different code entities like libraries, functions and classes. The method is also customizable as per end-user needs via the code entity weights. For instance, if the user wants demonstration examples to be more similar in terms of \textit{class} and not much in terms of \textit{functions} and \textit{libraries}, the importance can be adjusted by tuning the weight parameter $w_{e_{i}}$ suitably, where $e_{i}$ is a particular entity.  

\vspace{2mm}
\noindent
\textbf{RQ4 : How do the token-based and embedding-based strategies compare?}

We perform a comparative study between $Selection_{token}$, $SSL_{ner}$ (both token-based) and $Selection_{semantic}$ (embedding-based). 

For the CoNaLa dataset, we find the best performance is observed in StarCoder (ref. Table~\ref{tab:conala_quant}). The improvements over the best token-based method $SSL_{ner}$ and $Selection_{semnatic}$ are trivial and is not statistically significant.
Similar observations hold for the five intents in the TLC dataset (Tables~\ref{tab:tlc_use_quant} -- ~\ref{tab:tlc_what_quant}). 

We now look at a qualitative example from the TLC dataset (intent: ``use'') in Figure~\ref{fig:compare}.  Due to the lengthy function-level codes and page limitation, we omit portions of the selected codes in the middle (ref. ~\ref{sec:app1} for details). The query code has the ground truth ``invoked when an action from the popup menu occurs''. We show the top 1 example selected by each SSL-approach $Selection_{token}$, $Selection_{semantic}$ and $SSL_{ner}$ and the corresponding explanations of the query code generated by StarCoder for each demonstration example. 

For $Selecction_{token}$ we find that the explanation generation is not accurate and straight-forward. It is also difficult to understand the points of similarity between the demonstration example and the query code. $Selection_{semantic}$ gives a much better explanation of the query code compared to $Selection_{token}$ as it hints at some user clicks and action occurring thereafter. The reason behind the selection of this example is difficult to interpret as there are no direct links observable. For instance the query code uses methods like \texttt{getSource()} and classes like \texttt{OMGraphicHandler}. The example from $Selection_{semantic}$ consists of classes like \texttt{DefaultMutableTreeNode} and methods like \texttt{getRoot()}. For $SSL_{ner}$ we find the example consisting of similar methods \texttt{getSource()} and class \texttt{JMenuItem}. The explanation generated by the LLM using this demonstration example is hence similar to the ground truth explanation, although it misses the word ``popup'' .

\label{sec:results}

\section{Conclusion and Future Work}
\vspace{-1mm}
In this paper, we perform a comparative study of several open-source Code LLMs, SSL methods and experiment with two datasets having varying levels of explanations for the code explanation task. We perform a thorough analysis of the methods and the performances of the different CodeLLMs that lead to different interesting insights. To the best of our knowledge, this is also the first systematic attempt for evaluating open-source Code-LLMs for the task.

Additionally, we introduce a new Selective-shot Learning method $SSL_{ner}$ based on code-based NER . Empirical results suggest $SSL_{ner}$ to be the best token-based demonstration selection strategy 
while being inherently interpretable and customizable through the code entities. 

There are several avenues to extend this work. Possibilities of combining $SSL_{ner}$ with embeddings may be studied. We also plan to experiment with repository level code explanations. Fine-tuning the LLMs by using the relevant examples selected by $SSL_{ner}$ is likely to improve performance. We leave its consideration to future research.

\label{sec:concl}

\section{Limitations}

\noindent
In this work, we do not perform finetuning the models for the code explanation task and compare the results, since the focus of the paper is on the unsupervised evaluation of the LLMs for the task. We do not include ASAP~\cite{segment} in our study as it uses additional information like repository information, data flow graph, AST tree, variable names etc. The scope of our study is restricted to using $(code\ snippet, code\ explanation)$ pairs as exemplars.
\begin{table}[!thb]
\caption{Examples showing the limitation of $SSL_{ner}$}
\label{tab:drawback}
\resizebox{\linewidth}{!}{
\begin{tabular}{|c|l|l|l|}
\hline
\textbf{\begin{tabular}[c]{@{}c@{}}Change\\ ROUGE-L\end{tabular}} & \multicolumn{1}{c|}{\textbf{Reference}} & \multicolumn{1}{c|}{\textbf{$Selection_{token}$}} & \multicolumn{1}{c|}{\textbf{$SSL_{ner}$}} \\ \hline
-14\% & \begin{tabular}[c]{@{}l@{}}helper method to check \\ for equality  between two\\ object , including null checks .\end{tabular} & \begin{tabular}[c]{@{}l@{}}utility method for checking \\ equality between two objects, \\ coping with nulls\end{tabular} & \begin{tabular}[c]{@{}l@{}}helper method for checking \\ equality between two objects\end{tabular} \\ \hline
-33\% & \begin{tabular}[c]{@{}l@{}}add a view for the dummyview\\ to  draw .\end{tabular} & \begin{tabular}[c]{@{}l@{}}Add a fake view to\\ the layout.\end{tabular} & \begin{tabular}[c]{@{}l@{}}adds a fake view to the list\\ of  views that will be used to\\ calculate the height of the view .\end{tabular} \\ \hline
-59\% & \begin{tabular}[c]{@{}l@{}}computes the result for one \\ input double value .\end{tabular} & \begin{tabular}[c]{@{}l@{}}The code computes the\\ result for   a double input\\ value and returns a date.\end{tabular} & \begin{tabular}[c]{@{}l@{}}The code is a method named \\ `compute` that takes a double value\\ as input and returns a Date object.\end{tabular} \\ \hline
\end{tabular}
}

\end{table}
The limitations of $SSL_{ner}$ may be summarized as follows:~(i) The method incurs an overhead of running an code entity extraction module. Nevertheless, if the model is implemented in production, the entity extraction on the training examples can be run in batch mode and stored in a database. For the query code snippet, the entity extraction has to be done on the fly. ~(ii) The explanations given by $SSL_{ner}$ are more descriptive compared to the reference explanation which is more generic. Table~\ref{tab:drawback} shows a few examples where the performance of $SSL_{ner}$ is worse compared to $Selection_{token}$. ~(iii) There might be cases where the training dataset does not have code entities similar to a given code snippet. In such scenarios, the plausible approach is to switch to the $Selection_{token}$ strategy.

\label{threat}

\section{Ethical Considerations}
One significant issue when dealing with LLMs involves the risk of inadvertently exposing test data during the training process. Unfortunately, it is not feasible to directly verify this as the training dataset remains inaccessible. The model's relatively lower performance with zero-shot and random few-shot techniques indicates that memorization might not play a significant role. As we introduce more relevant and contextual information, the model's performance gradually enhances with the quality of the provided data. If the model had already memorized the summaries of the test code snippets beforehand, it could have achieved substantially higher scores, even in a zero-shot setting. One way to alleviate this concern would be to train the models from scratch. However, this would incur high computational costs and is currently infeasible.\\

We did not perform a validation of the code entities predicted by the UniversalNER model. Since this module forms the backbone of the proposed approach, an evaluation is necessary. We leave this as a future work.
We have not carried out a user study for our approach. Therefore, the improvements in the evaluation metrics presented in the paper may not directly correlate with improvement in end-user satisfaction. This aspect remains to be explored as a future research.
\label{ethics}

\bibliography{custom}
\bibliographystyle{acl_natbib}

\appendix

\section{Appendix}
\label{sec:appendix}
\subsection{Qualitative Example for Comparing the three SSL approaches}
\label{sec:app1}
\begin{figure*}[!thb]
\caption{An example demonstrating the Query Code method, the top 1 demonstration example selected by $Selection_{token}$, $Selection_{semantic}$ and $SSL_{ner}$ along with the LLM (StarCoder) generated output for each method, respectively. Due to the lengthy function-level codes and page limitation, we omit portions of the selected codes in the middle.}
\label{fig:compare_app}
\centering
\fbox{
\includegraphics[width=16cm]{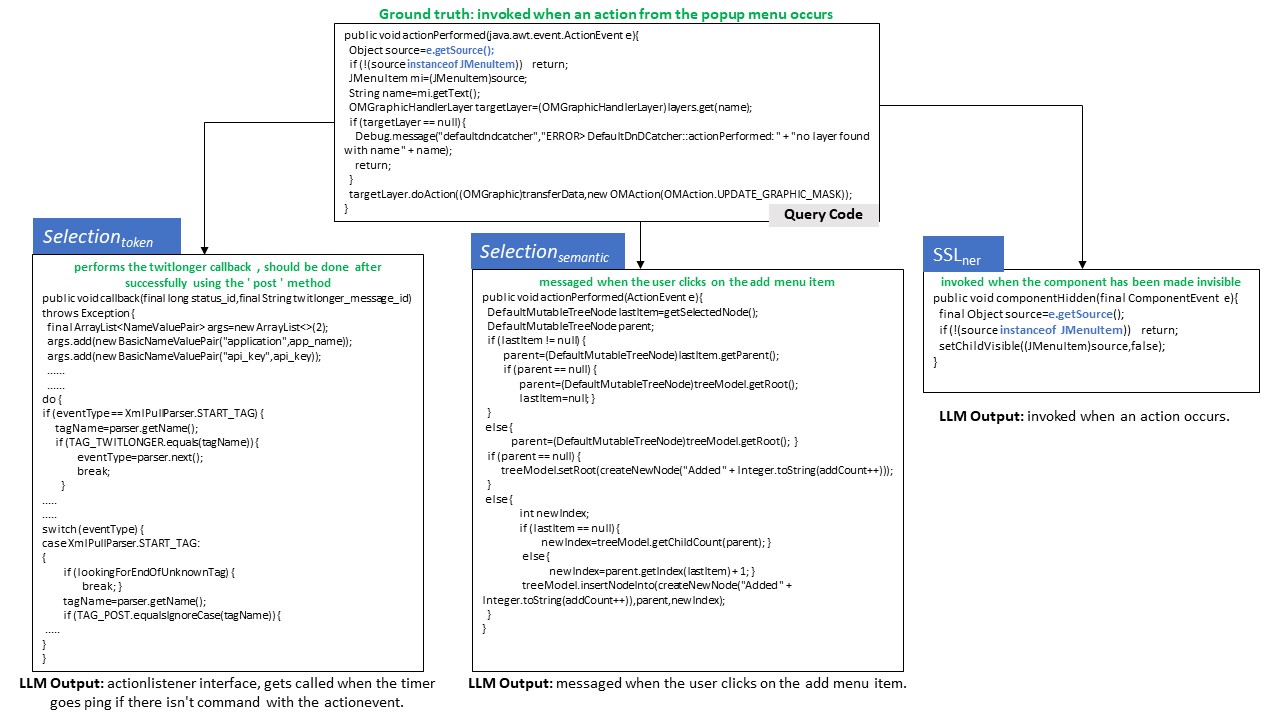}
}
\vspace{-7mm}
\end{figure*}
This is a more detailed analysis of RQ4 in Section~\ref{sec:results} with Figure~\ref{fig:compare_app} containing a more detailed view of the code snippets. We repeat the analysis here, for ease of readability.

The qualitative examples are from the TLC dataset (intent: ``use''). The query code has the ground truth ``invoked when an action from the popup menu occurs''. We show the top 1 example selected by each of the demonstration selection strategies $Selection_{token}$, $Selection_{semantic}$ and $SSL_{ner}$ and the corresponding explanations of the query code generated by StarCoder for each demonstration example. 

For $Selecction_{token}$ we find that the explanation generation is not accurate and straight-forward. It is also difficult to understand the points of similarity between the demonstration example and the query code. $Selection_{semantic}$ gives a much better explanation of the query code compared to $Selection_{token}$ as it hints at some user clicks and action occurring thereafter. The reason behind the selection of this example is difficult to interpret as there are no direct links observable. For instance the query code uses methods like \texttt{getSource()} and classes like \texttt{OMGraphicHandler}. The example from $Selection_{semantic}$ consists of classes like \texttt{DefaultMutableTreeNode} and methods like \texttt{getRoot()}. For $SSL_{ner}$ we find the example consisting of similar methods \texttt{getSource()} and class \texttt{JMenuItem}. The explanation generated by the LLM using this demonstration example is hence similar to the ground truth explanation, although it misses the word ``popup'' .

\subsection{Qualitative Examples for Comparing the Token-based and Code-Entity based Methods}
\label{sec:app2}
\begin{figure*}[!thb]
\caption{Figure demonstrating a query code sample, the top 3 examples selected by $Selection_{token}$ and the explanation generated for the query code sample using StarCoder. Due to the lengthy function-level codes and page limitation, we omit portions of the selected codes in the middle.}
\label{fig:cluster_token}
\centering
\fbox{
\includegraphics[width=16cm]{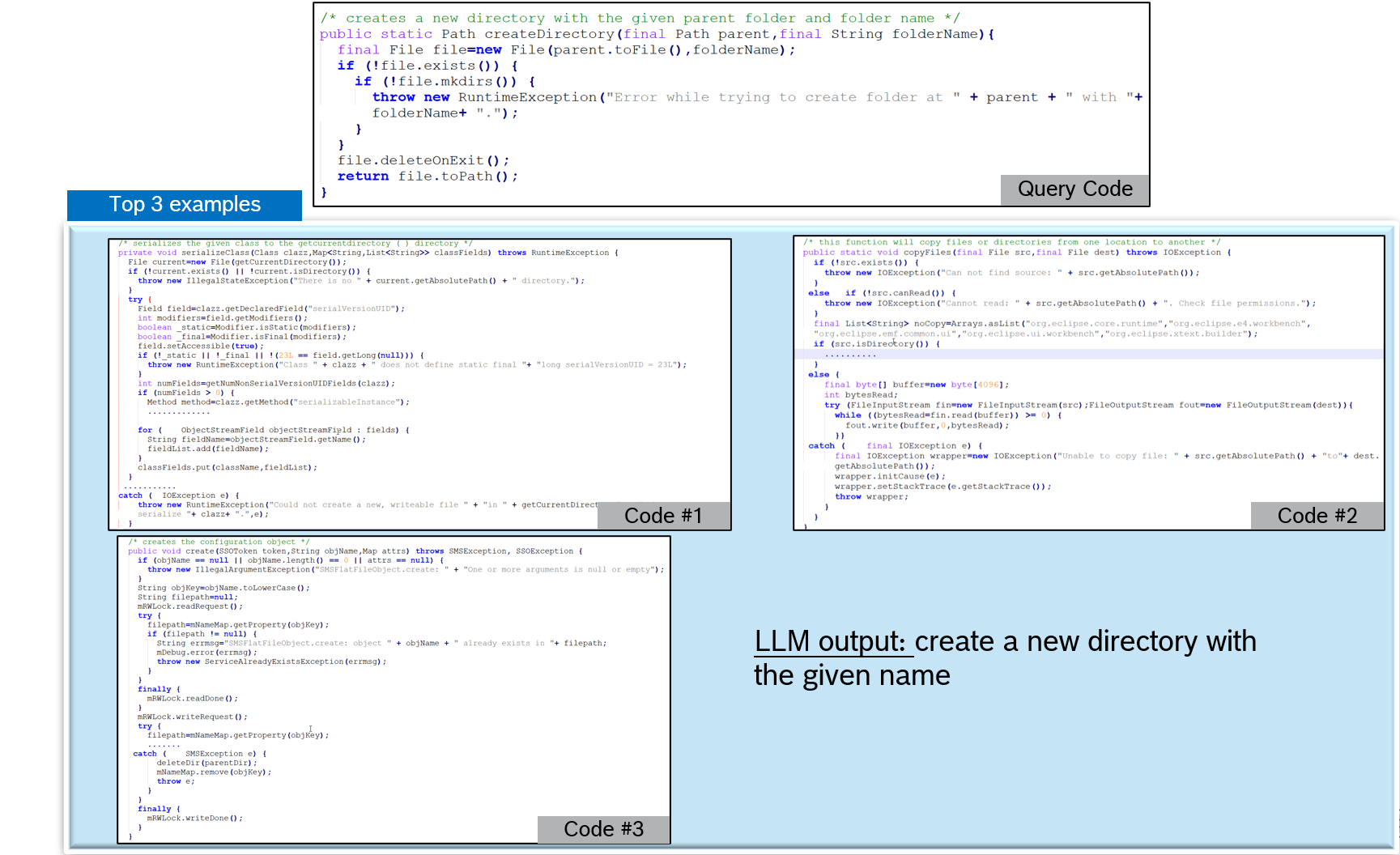}
}
\end{figure*}

\begin{figure*}[!thb]
\caption{Figure demonstrating a query code sample and the top 3 examples selected by $SSL_{ner}$ and the explanation generated for the query code sample using StarCoder.}
\label{fig:cluster_ner}
\centering
\fbox{
\includegraphics[width=16cm]{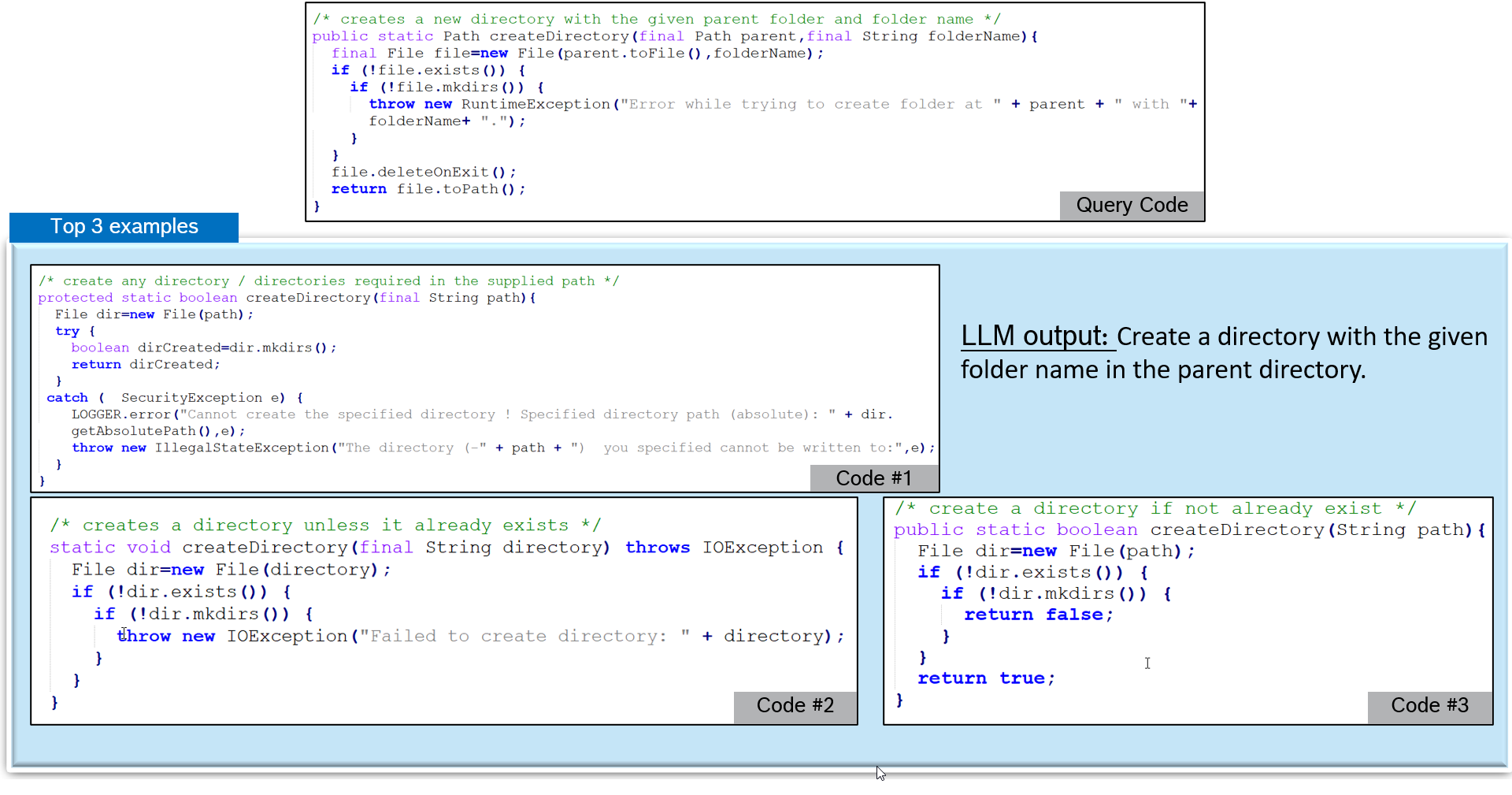}
}
\end{figure*}

Figures~\ref{fig:cluster_token} and ~\ref{fig:cluster_ner} shows the top-3 examples selected by $Selection_{token}$ and $SSL_{ner}$ respectively given the query code. The ground truth explanation for the query code is ``creates a new directory with the given parent folder and folder name''.

As can be observed, the query code is on creating a new directory with the given parent folder and folder name. The top 3 examples selected by $Selection_{token}$ are based on serializing a class, copying files from one location to another and creating a configuration object. From a qualitative perspective it is difficult to figure out why the method has selected the codes, as one cannot directly observe similar classes, functions or methods with respect to the query code. The explanation that is generated for the query code using StarCoder, ``create a new directory with the given name'' is not accurate as it misses the location as to where the new directory is to e created.

$SSL_{ner}$ on the other hand, selects codes that are directly on creating directories, but has different variations on handling the runtime errors. The similarity notion arises from the fact that the query code and the top 3 code examples use the same functions (mkdirs, exists) and class (File). Hence these examples are more relevant to the query code. The explanation that is generated for the query code using StarCoder, ``create a directory with the given folder name in the parent directory'', is also more accurate.

Hence we  conclude that $SSL_{ner}$ selects more relevant demonstrations that can be interpreted through the matches in different code entities like libraries, functions and classes. $Selection_{token}$ although a more simpler approach, lacks this interpretability.

\end{document}